\documentstyle[sprocl]{article}
\newcommand{\newc}{\newcommand}

\newc{\gsim}{\lower.7ex\hbox{$\;\stackrel{\textstyle>}{\sim}\;$}}
\newc{\lsim}{\lower.7ex\hbox{$\;\stackrel{\textstyle<}{\sim}\;$}}

\def\beq{\begin{equation}}
\def\eeq{\end{equation}}
\def\beqn{\begin{eqnarray}}
\def\eeqn{\end{eqnarray}}

\def\inbar{\,\vrule height1.5ex width.4pt depth0pt}

\def\IC{\relax\hbox{$\inbar\kern-.3em{\rm C}$}}
\def\IQ{\relax\hbox{$\inbar\kern-.3em{\rm Q}$}}
\def\IR{\relax{\rm I\kern-.18em R}}
 \font\cmss=cmss10 \font\cmsss=cmss10 at 7pt
\def\IZ{\relax\ifmmode\mathchoice
 {\hbox{\cmss Z\kern-.4em Z}}{\hbox{\cmss Z\kern-.4em Z}}
 {\lower.9pt\hbox{\cmsss Z\kern-.4em Z}}
 {\lower1.2pt\hbox{\cmsss Z\kern-.4em Z}}\else{\cmss Z\kern-.4em Z}\fi}
\def\NPB#1#2#3{{\it Nucl.\ Phys.}\/ {\bf B#1} (19#2) #3}
\def\PLB#1#2#3{{\it Phys.\ Lett.}\/ {\bf B#1} (19#2) #3}
\def\PRD#1#2#3{{\it Phys.\ Rev.}\/ {\bf D#1} (19#2) #3}

\def\beq{\begin{equation}}
\def\eeq{\end{equation}}
\def\beqn{\begin{eqnarray}}
\def\eeqn{\end{eqnarray}}

\begin{document}
\rightline{\tt hep-ph/0005033}
\rightline{DAMTP-43-2000} 
\bigskip

\title{ON REALISTIC BRANE WORLDS FROM TYPE I STRINGS\footnote{Plenary
talk by FQ at PASCOS 99}}
\author{Gerardo Aldazabal}

\address{Instituto Balseiro, CNEA, Centro At\'omico Bariloche,\\[-0.3em]
8400 S.C. de Bariloche, and CONICET, Argentina\\E-mail:
 aldazaba@cab.cnea.gov.ar}

\author{Luis E. Ib\'a\~nez}

\address{ Departamento de F\'{\i}sica Te\'orica C-XI
and Instituto de F\'{\i}sica Te\'orica  C-XVI,\\[-0.3em]
Universidad Aut\'onoma de Madrid,
Cantoblanco, 28049 Madrid, Spain\\
E-mail: ibanez@madriz1.ft.uam.es}  

\author{Fernando Quevedo}

\address{DAMTP, CMS Wilberforce Road, Cambridge CB3 0WA, UK\\
E-mail:f.quevedo@damtp.cam.ac.uk}


\maketitle

\abstracts{
We review recent progress in constructing realistic brane models from type I
string vacua. Explicit models with three families of the standard
model gauge group and its left-right generalizations are presented
with supersymmetry broken at the string scale of order $M_s\sim
10^{10-12}$
 GeV,
realizing gravity mediated supersymmetry breaking at low energies.
Unification of couplings occurs
 at the string scale due to the
particular $U(1)$ normalizations of D-branes,
 as well as to the existence of a Higgs
field per family of quarks and leptons. The proton is naturally stable
due to intrinsic discrete symmetries of the corresponding string
theory. In particular $R$-parity appears as a natural stringy
symmetry.
There are  axionic fields with the right couplings as
to solve the strong CP problem. Similar realizations are also
presented for a string scale of $1$ TeV, although without solving 
the gauge unification problem. Open questions 
are briefly discussed.}

\section{Introduction}

The present understanding of string theory indicates that all the
different
10-dimensional string theories (types I, IIA, IIB and two heterotic)
 happen to be different manifestations of a single $M$-
theory. It has also led to a prime role played by high dimensional
surfaces known as D-branes, giving support to the idea that our
4-dimensional 
world could  itself be  a brane.

The brane world scenario has been subject to intense investigation
during the past two years and new interesting mechanisms have been
proposed to solve longstanding problems with the standard model, 
such as the hierarchy problem, gauge coupling unification, neutrino
masses, strong CP problem, etc. 
One of the interesting properties of this scenario is that it allows
for a fundamental scale of nature to be much below the Planck scale
and therefore closer to experiments \cite{here} .
However until recently  explicit realizations of this
scenario, with low-energy fundamental scale, were lacking. We review
here the progress we have made  in that direction during the past few months.
 
Since this is the first talk of the meeting we have to briefly review
the idea of the brane world scenario. This is a variation of a
Kaluza-Klein theory for which the extra dimensions are felt only by a
subset of the fields. The typical case is that the Standard Model
fields are constrained to live inside a low dimensional 
surface, or brane, of the high
dimensional spacetime, but gravity lives in the full spacetime
\cite{joe}. 
This seems like a {\it ad-hoc} separation of the fields, however
recent developments on string theory precisely point at this scenario.
The Horava-Witten realization of strongly coupled heterotic string,
leads after compactification, to a 5D world with the 5th coordinate
being just an interval. Gauge and matter fields live only on 
the two 4D surfaces at each end of the
interval, whereas gravity lives in the full 5D spacetime. Similarly,
and more relevant for this talk, type I string theory includes
Dp-branes.
These are surfaces where the end points of the open strings are
attached,
satisfying Dirichlet boundary conditions, here the origin of the name 
D-branes.
The surfaces may be of different dimensionality which is denoted by
$p$. Each Dp-brane has a $U(1)$ gauge field corresponding to an open
string with both endpoints on the same brane. There are also states
corresponding to open strings with endpoints on two different branes,
the mass of these states is proportional to the distance between the
branes, therefore when two branes overlap, the distance vanishes and
these states become  massless with the net effect of enhancing the
gauge symmetry from $U(1)^2$ to $U(2)$. If there are $N$ overlapping
branes the gauge symmetry will then be $U(N)$. On the other hand,
gravity corresponds to closed strings and these can move on the whole
10-dimensional spacetime.
 Therefore we have a clean realisation of the brane world
scenario in type I string theory with $U(N)$ (or other groups) living
on the brane and gravity on the bulk.

\section{Brane World versus Kaluza-Klein}

It is important to realise the difference between the brane world and
the better known Kaluza-Klein scenario. In Kaluza-Klein all fields
feel the extra dimensions whereas in the brane world, only a subset of
the fields (gravity and moduli fields in string theory) feel all the
extra dimensions. 

This simple fact has very important physical implications regarding
the possible values of the fundamental scale. An explicit way to see
the difference is comparing the low-energy effective actions for 
perturbative heterotic strings and type I strings. In the heterotic
case, both gravity and the gauge fields live on the full
10-dimensional spacetime corresponding to a standard Kaluza-Klein
scenario. The low-energy effective action in 10 dimensions takes the
form:
\begin{equation}
S\ = \ M^8\  \int d^{10}x\sqrt{-G}\ e^{-\phi}\left( {\cal R}\ + \ M^{-2}\
F_{MN}^2
\ +\ \cdots \right),
\end{equation}
where $M=1/\sqrt{\alpha'}$ is the  string scale and $\phi$ is
the dilaton field.
Upon compactification to 4-dimensions each of the two terms in the
action above will get a volume factor coming from the integration of 
the 6 extra dimensions. This gives us an expression for the
gravitational and gauge couplings (the numerical coefficients of each
of the two terms above) of the form:
\begin{equation}
M_{Planck}^2\ \sim \ e^{-\phi}\ M^8 r^6\qquad \alpha_{GUT}^{-1}\ 
\sim\ e^{-\phi}M^6 r^6
\end{equation}
where $r$ is the overall size of the extra dimensions.
Taking the ratio of those expressions the volume factors cancel
and we get $M_{Planck}^2\sim \alpha_{GUT}^{-1} M^2$. Therefore for 
$\alpha_{GUT}$ not much different from 1 (as expected) we have to have
the fundamental scale $M$ to be of the same order of magnitude as the 
gravitational scale $M_{Planck}\sim 10^{19}$ GeV. This was the old
belief that the string scale was  the Planck scale.

Things are very different in the brane world scenario as we can see
for the case of the type I string.
For a configuration with the standard model spectrum belonging to 
a Dp-brane, the low energy action in 4-dimensions takes the form:
\begin{equation}
S\ =\ -\frac{1}{2\pi} \int d^4x\sqrt{-g}\left({r^6\
M^8}{\ e^{-2\phi}}{\cal R}\ +\ \frac{1}{4}{(rM)^{p-3}}{\ e^{-\phi}}\ F_{\mu\nu}^2\
+\ \cdots \right)
\end{equation}
Comparing the coefficient of the Einstein term with the physical
Planck mass $M_{Planck}^2$ and the coefficient of the gauge kinetic
term with the physical gauge coupling constant $\alpha_p$($\sim 1/24$
at the string scale), we find the relation:
\begin{equation}
 M^{7-p}\ =\ \frac{\alpha_p}{\sqrt 2}\ M_{Planck}\ r^{p-6}
\end{equation}
from which we can easily see that if the Standard Model
fits inside a D3-brane, for instance, 
 we may have $M$ substantially smaller than
$M_{Planck}$ as long as the sizes of the extra dimensions are large enough.

Given the fact that we do not have a way to fix the size of the extra
dimensions we can take advantage of our ignorance 
and  follow a bottom-up approach  considering different 
possibilities motivated by phenomenological inputs.
Several scenarios have been proposed depending on the value of the
fundamental scale. The four main scenarios at present correspond to 
\begin{enumerate}
\item{} $M\sim M_{Planck}$. This is just the old perturbative heterotic
string case
corresponding to compactification scale close to the Planck scale. 
There is nothing wrong with this possibility
\cite{faraggi}. Research over the years has shown
difficult to obtain gauge coupling unification in this case.

\item{} $M\sim M_{GUT}\sim 10^{16}$ GeV. Obtained for $r\sim 10^{-30}
cm$ in the expression above. This proposal \cite{witten}\  was
made precisely to `solve' the gauge coupling unification problem in
string theories. This requires a compactification scale of order
$10^{14}$ GeV. 
Recent progress has been made \cite{ovrut} in looking for three
generation models  realising this
scenario from the Horava-Witten construction
 but, so far,  not from
type I models.

\item{} $M\sim M_I\sim 10^{10-12}$ GeV. If the world is a D3-brane we
can see that this scale is obtained from the equations above for 
$r\sim 10^{-23}cm$. This proposal
\cite{benakli}, was based on the special role played by the
intermediate scale $M_I$ in different ideas beyond he standard model. 
Particularly the scale of supersymmetry breaking in gravity mediated
supersymmetry breaking scenario. This then allows to identify the
string scale with the supersymmetry breaking scale and opens up the
room for  non supersymmetric string models to be relevant at
low-energies. Explicit models realising this scenario will be discussed
in the next section.

\item{} $M\sim M_{EW}\sim 10^3$ GeV.  This is obtained for overall
radius $r\sim 10^{-12}cm$ above and if only two of the six dimensions 
were large this would have given us the famous $r\sim 1 mm$ quoted as
the extreme case of the brane world scenario 
since lengths bigger than this
would have been observed by deviations of gravity.
 This is the most popular scenario \cite{joe2} 
due to its proximity with experiment. Concrete string models realising
this scenario do not exist. We will discuss some 
attempts in the next section.

\end{enumerate}

Notice that only the first scenario was possible following the
standard Kaluza-Klein approach in the perturbative heterotic string
models.
The brane world  opened up the possibility of the next three
scenarios as well as any other scale in the range
$M_{EW}<M<M_{Planck}$.

\section{String Model Building}
In order to construct realistic string models, starting from 10 (or
11) dimensions we have to specify the nature of the extra 6 or 7
dimensions.
The simplest idea that comes to our mind is to set the extra
dimensions to be circles or tori, since they are still flat, however
the 4-dimensional models obtained in this way are not chiral for any 
string theory and therefore cannot describe the standard model.

In the 80's this was solved by either assuming that the extra dimensions
were a compact Calabi-Yau space or an orbifold limit
\cite{rev}.
 An orbifold is a twisted torus. The simplest
example is the one-dimensional interval which can be considered as a
twisted circle in the following way: defining the circle as the set of
all
the points in the real line identified by a  $2\pi r$ shift, 
{\it ie} $X\equiv\ X+2\pi r$, we can now twist the circle by further
identifying
the points $X\equiv -X$, which leaves us only with the interval
$[0,\pi r]$. The two end points of the interval are fixed points under
the orbifold identification and represent `singular' points of the
space. This is then a $Z_2$ orbifold in one dimension.
A generalization to 6 dimensions in terms of more general discrete
twists
defines  good compactifications of
string theory.

There are many possible string vacua constructed this way, they are
determined
by the different allowed twists of six-dimensional tori, the way this
twists acts on the gauge degrees of freedom and the addition of
nontrivial
gauge background fields on the noncontractible loops of the defining
tori.
These `Wilson lines' are the main source for the huge degeneracy of this
class of string vacua\cite{rev}.

Explicit orbifold models were constructed in the past starting from
the heterotic string, exhibiting many
features similar to the standard model (same gauge group, three
families,
structure of Yukawa couplings, etc.)\cite{rev}
 and new constructions are still
being worked out \cite{faraggi}.
The orbifold construction has the advantage over Calabi-Yau manifolds
of being essentially a
flat space, except for a few special points and still allowing chiral
4-dimensional models and therefore it is possible to describe string
interactions in such a background.
It is worth mentioning that these compactifications singled out the
heterotic string as the most viable string theory since it was the one
for which it was possible to obtain chiral models with realistic properties.

\section{Realistic Type I Brane Models with Broken SUSY}

After the introduction of D-branes, the perspective about type I
models has changed completely. It allows us to look for the standard
model not only inside the `bulk' 10-dimensional spacetime but also
inside
some of the lower dimensional D-branes that appear in such vacua.

Much work has been devoted recently to the construction of
4-dimensional type I models. A particularly useful way to build open type I
string models is to start with closed type IIB strings and perform a
kind of orbifold twist on the (2-dimensional) string worldsheet
identifying the two orientations of type IIB strings, this is called
an `orientifold'. On top of this, compactifications similar to those of the
heterotic string in terms of orbifolds have been obtained, classified
by the different twists and background gauge fields or Wilson lines.
The net result of this investigation is that although similar to the
heterotic strings, many chiral models can be constructed preserving
$N=1$ supersymmetry, none of them can be claimed to be close to the
Standard Model.

One of the reasons for this lack of realistic models is the fact that
there are consistency conditions that the models have to satisfy in
order to avoid unwanted tadpoles (which if existing would give rise to
anomalies in the 4-dimensional theory). These tadpole cancellation
conditions happen to be more restrictive than the corresponding
conditions in the heterotic case and therefore there is less room for
realistic models.

In order to obtain realistic models we may relax the conditions we had
imposed on the models, in particular we may look for models without
supersymmetry. Notice that this possibility was not open to us in the
heterotic case because constructing a nonsupersymmetric model at the
Planck  scale would leave us without a solution to the hierarchy problem.
As argued in the previous section, in type I models we may have the
string scale lower than $10^{12}$ GeV. In this case having a
nonsupersymmetric model may still solve the hierarchy problem at low
energies as long as we have gravity mediated supersymmetry breaking in
the visible sector of scenario 3. above, 
 for which the splitting in multilplets will be of
the order $M^2/M_{Planck}\leq 1$ TeV.
On scenario 4. we may just have explicit supersymmetry breaking
without any danger. Therefore in nonsupersymmetric brane models are
now
an interesting alternative to the supersymmetric string vacua.

A concrete way to build nonsupersymmetric brane models is to look for
string vacua including both branes and antibranes. It is known that 
a brane, being a BPS state, breaks partially supersymmetry, an
anti-brane breaks the remaining supersymmetry so the configuration
brane/anti-brane is non supersymmetric.
However brane/anti-brane configurations tend to be unstable which
usually shows in the appearance of tachyons in the spectrum. In 
orbifolds of type I models this can be avoided by having the antibranes
of different dimension than the branes which then do not annihilate
each other. Furthermore, tadpole cancellation conditions force some of
the branes or anti-branes to be trapped in some of the orbifold fixed
points
avoiding the annihilation of branes and anti-branes of the same
dimensionality.

We can then envisage models with,for instance, D7-branes with
D3-branes trapped at some of the orbifold fixed points and
some anti D3-branes trapped at different orbifold singularities
which cannot annihilate each other.
Models of this type have been explicitly constructed recently
\cite{aiq1,aiq2} with the following physical properties:
\begin{enumerate}
\item{}On the D7-branes there is the gauge symmetry $SU(3)_c\times
SU(2)_L\times U(1)_Y$ or its left-right extension $SU(3)_c\times
SU(2)_L\times SU(2)_R\times U(1)_{B-L}$ the matter sector includes
three families of quarks and Higgs fields.
\item{}On the trapped D3-branes and anti D3-branes there are extra
gauge
fields which for the D3 case are broken by some flat directions in the
model. The three families of lepton fields appear as open strings with 
one endpoint on a D3-brane and the other on the D7-brane. 
\item{}The presence of `hidden' anti-D3-branes explicitly break
supersymmetry,
but its breaking is felt by the visible sector only through
gravitational
strength interactions. Therefore if the string scale is the
intermediate scale this would correspond to the gravity mediated
supersymmetry breaking scenario. If the anti-D3 branes happen to be
inside the D7-branes then the breaking of supersymmetry is explicit
and a  TeV fundamental scale is required.
\item{}The D-brane origin of the $U(1)$ gauge groups fixes the
hypercharge normalization to be $3/14$ different from the $3/8$ of
$SO(10)$
GUTs. Furthermore the appearance of three rather than one families of
Higgs fields change the RG running of the couplings in such a way that
unification occurs at the intermediate scale $M\sim 10^{11}$ GeV. This
is particularly realised in the left-right models for the scale of 
$SU(2)_R$ breaking close to 1TeV, having then important experimental
consequences at present and future collider. The 1TeV scenario fails
to satisfy the gauge coupling unification in this class of models.
\item{} Yukawa couplings providing structure of quarks and lepton
masses
(including neutrino masses) can be obtained from the 
 the superpotential. Their full understanding requires also knowledge of
the Kahler potential which is not  under complete control.
\item{}The particular way that the three lepton families are
distributed among the branes gives rise to discrete versions of lepton
number as exact symmetry of the models. Further discrete symmetries
obtained from the structure of the original flat directions, give rise
naturally to $R$-parity as an exact discrete symmetry forbidding fast
proton decay. 
\item{}There are particular (Ramond-Ramond) axion fields at the
singularities with the properties and couplings needed to solve the
strong CP problem. This depends crucially on the fact 
that the fundamental scale
is intermediate. This was not possible in the old heterotic models. 
\item{}The explicit potential for the scalar fields is not known and
the minimisation process cannot be performed at present. It is yet
to be seen that this stabilisation could fix the value of the compact
space to the `right value' as to obtain the intermediate fundamental
scale
and the even most difficult requirement of generating a very small
cosmological constant which is not protected after supersymmetry was
broken. These are left as open questions for the moment.
\end{enumerate}
 These are exciting times for string model building.


\begin{thebibliography}{99}
\bibitem{joe} For a mechanism where gravity is also localised
 see the
contributions of Lykken and Csaki and references therein.
\end{thebibliography}
\end{document}